\documentclass[a4paper,11pt]{article}
\pdfoutput=1 
 
\usepackage{jcappub} 
                      
\usepackage[T1]{fontenc} 
\usepackage{graphicx} 
 
\title{\boldmath Extending the Coyote emulator to dark energy models 
  with standard $w_0$-$w_a$ parametrization of the equation of state.} 
 
\author[a,1]{L. Casarini,\note{Corresponding author.}} 
\author[b]{S.A. Bonometto} 
\author[b]{E. Tessarotto} 
\author[c]{and P.-S. Corasaniti} 
 
\affiliation[a]{Federal University of Espirito Santo (UFES), Physics  
Department, Vitoria ES, Brazil} 
\affiliation[b]{INAF, Observatory of Trieste \& Trieste University, 
  Physics Dep., Astronomy Unit, Via Tiepolo 11, 34143 Trieste, Italy} 
\affiliation[c]{LUTH, Observatoire de Paris, PSL research University, 
  CNRS, Universit\'e Paris Diderot, Sorbonne Paris Cit\'e, 5 Place 
  Jules Janssen, 92195 Meudon, France} 
 
\emailAdd{casarini.astro@gmail.com} 
 
\abstract{We discuss an extension of the Coyote emulator to predict 
  non-linear matter power spectra of dark energy (DE) models with a 
  scale factor dependent equation of state of the form 
  $w=w_0+(1-a)w_a$. The extension is based on the mapping rule between 
  non-linear spectra of DE models with constant equation of state and 
  those with time varying one originally introduced in 
  ref.~\cite{Casarini2009}.  Using a series of N-body simulations we 
  show that the spectral equivalence is accurate to sub-percent level 
  across the same range of modes and redshift covered by the Coyote 
  suite. Thus, the extended emulator provides a very efficient and 
  accurate tool to predict non-linear power spectra for DE models with 
  $w_0$-$w_a$ parametrization. According to the same criteria we have 
  developed a numerical code that we have implemented in a dedicated 
  module for the CAMB code, that can be used in combination with the 
  Coyote Emulator in likelihood analyses of non-linear matter power 
  spectrum measurements.  All codes can be found at 
  \url{https://github.com/luciano-casarini/pkequal}.} 
   
\begin{document} 
\maketitle 

\section{Introduction}\label{intro}  
The origin of cosmic acceleration is scarcely understood. No doubt 
however remains that the Universe is close to spatially flat 
(curvature parameter $|\Omega_K| \lesssim 0.005$), while baryons and 
Dark Matter (DM) only account for $\sim 30\, \%$ of the critical 
density, errors hardly exceeding $2\, \%$ \cite{Planck_Cosmo}. 
Accordingly, we infere the gap to be filled by a component, or 
phenomenon, called Dark Energy (DE) and responsible for the observed 
accelerated expansion \cite{Riess1998,Perlmutter1999}. 
 
DE could be Einstein's cosmological constant $\Lambda$ and/or arise 
from vacuum quantum fluctuations, being so equivalent to a fluid with 
state parameter $w=-1$. This simple $\Lambda$CDM scenario meets a 
plenty of cosmic data, well beyond background composition, at least 
above middle size galactic mass scales. But this does not allow us to 
forget the mess of conceptual problems going with it, first of all 
that quantum field theory predicts a vacuum energy exceeding the 
phenomenological $\Lambda$ value by orders of magnitudes. 
 
An alternative to $\Lambda$ is a self--interacting scalar field $\Phi$ 
\cite{RatraPeebles1988,Wetterich1988,Ellis1989,CaldwellSteinhard1999}, 
possibly interacting with DM, as well. Moreover, General Relativity 
(GR) has no independent test on the huge scales where acceleration is 
measured, so DE could just arise from large scale GR violations (for 
comprehensive review see, e.g.,\cite{Clifton2012}). 
 
\begin{figure}[th]\label{fig:planckfig} 
\centering  
\includegraphics[width=.6\textwidth]{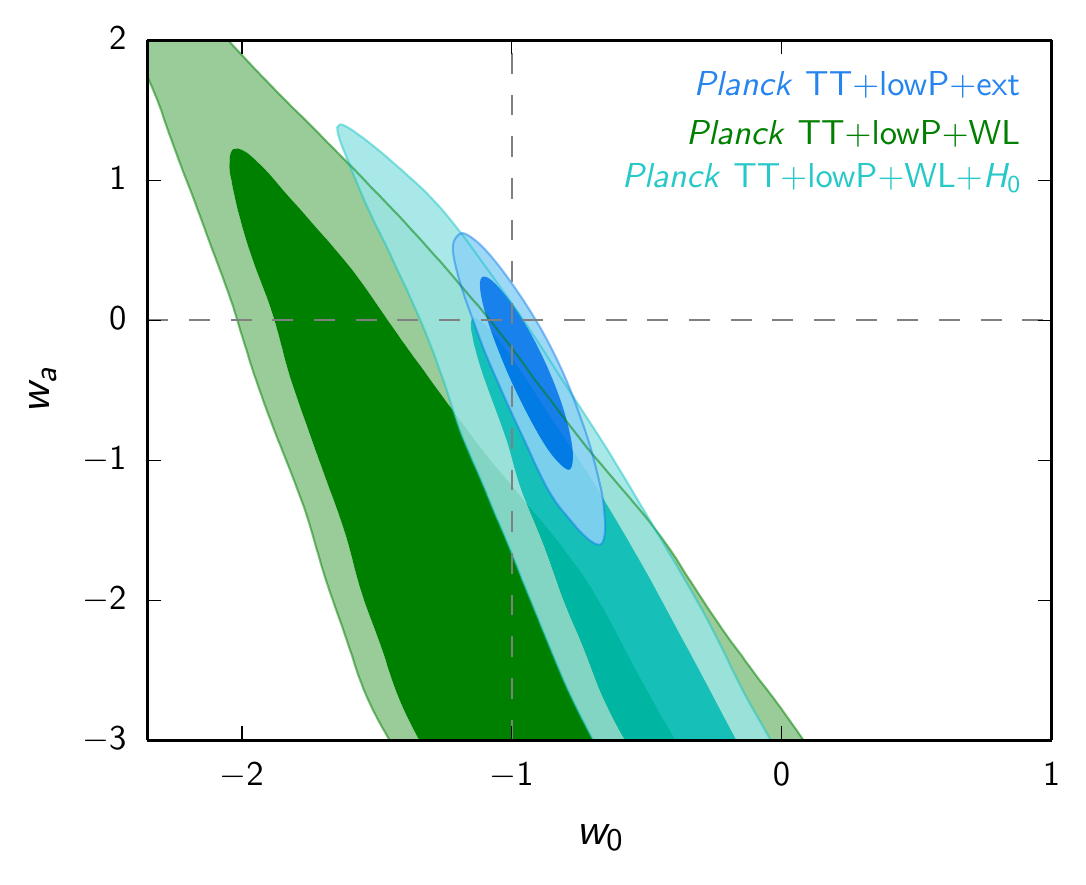} 
\caption{1 and 2$\sigma$ confidence contours in $w_0$-$w_a$ plane for 
  different combinations of cosmological datasets from 
  \cite{Planck_Cosmo} (courtesy of Planck collaboration).} 
\end{figure} 
Selecting among different options requires fresh data which, however, 
are hard to collect. The Planck experiment suggests $-1.10 < w < 
-0.94$ at $1\sigma$ \cite{Planck_Cosmo}, if $w$ is assumed to be 
constant. If however one assumes DE state parameter to be 
consistent with the expression \cite{Chevalier2000,Linder2003} 
\begin{equation} 
w(a)=w_0+(1-a)w_a 
\label{w0wa} 
\end{equation} 
and tests it against the same data, it turns out that $-1.25 < w_0 
<-0.65$ at $1\sigma$, as also shown by the small ellypse at the center 
of Figure \ref{fig:planckfig}.  (We will refer to eq.~(\ref{w0wa}) as 
CPL parametrization.)  Clearly results still significantly depend on 
the range of models tested.   
 
As shown by \cite{Bassett2004}, the CPL expression (and alike) is not 
exempt of pitfalls either. Moreover, Figure \ref{fig:planckfig} 
indicates that some tension exists between Planck outputs and 
CFHTLensS weak lensing data 
\footnote{Canada France Haway Telescope Lensing Survey} 
\cite{Erbenetal}, leadind to include values $|w_a| \gg 1$ in the 
likelihood ellypse, for which a CPL parametrization [eq.~(\ref{w0wa})] 
looses significance \cite{Corasaniti2002}. 
 
Although it is licit to suspect that some systematics pollutes such 
CMB--lensing comparison, these outputs make evident that weak lensing 
can be a basic tool to improve our DE knowledge. In fact, while CMB 
measurement errors have approached cosmic variance limits, so that 
planned future experiments are essentially meant to improve our 
knowledge on the inflationary era, the upcoming generation of 
tomographic lensing surveys, such as 
LSST\footnote{http://www.lsst.org/lsst/} and the {\sc Euclid} 
experiment \cite{laureijs}, can be expected to measure the power 
spectrum of matter density fluctuations, $P(k)$, across a wide range 
of scales and redshifts, up to $\sim 1\, \%$ accuracy. Such accuracy 
level is therefore the natural target for spectral predictions. 
 
On the very large length scales linear codes match such requirement. 
Perturbation theory can then extend predictions to the mild 
non--linearity area; in fact, aiming at a per-cent level, even on the 
scales of BAO we cannot neglect non-linear dynamics (see 
e.g. \cite{Rasera2014}). 

At larger $k$'s, only N-body simulations can provide model 
predictions. They however~demand a significant effort to overcome 
ambiguities due to sample variance, dependence on the box size and 
long-wavelength mode contributions, resolution and dynamical range 
across the entire model parameter space. In turn, this excludes N-body 
simulations to be part of an ``on line'' algorithm evaluating model 
likelihood. Accordingly, future galaxy survey data analyses require 
reliable and readily available spectral predictions from N--body 
simulations. 
 
This is why emulators are built, exploring a grid of models by varying 
a set of parameters, and based on wide set of simulations with 
different box sizes, resolution etc., for each grid element, so to 
minimize the impact of the overmentioned N--body simulation 
problems.  
 
The Coyote emulator  
\cite{Heitmann2014,Heitmann2010,Heitmann2009,Lawrence2010},  
in particular, is based on a suite of spatially 
flat simulations and provides 1$\, \%$ accuracy level estimates of 
spectra up to $k = 2\, {\rm Mpc}^{-1}h$ (at greater $k$ baryon physics 
matters more than 1$\, \%$), for the following parameter range 
$$ 
\begin{matrix} 
w~(const.) & ~~~~~ & \omega_b & ~~~~~ & \omega_m & ~~~~~ & n_s & ~~~~~  
& \sigma_8 \cr 
-1.3/-0.7  & ~~~~~ & 0.0215/0.0235 & ~~~~~ & 0.120/0.155 & ~~~~~ & 0.85/1.05 
& ~~~~~ & 0.6/0.9 \cr  
\end{matrix} 
$$ 
Here $\omega_{b,m}=\Omega_{b,m}h^2$, while $h$ and $\Omega_{b,m}$ are 
the dimensionless Hubble parameter and the density parameters for 
baryons and DM, respectively; $n_s$ is the spectral index for primeval 
scalar perturbations; $\sigma_8$ is the m.s. amplitude of density 
fluctuations on the 8$\, h^{-1}$Mpc scale. For a given value of $w$ 
and $\omega_{b,m}$ the reduced Hubble constant $h$ is set to reproduce 
the location of the CMB peaks in the temperature anisotropy power 
spectrum. The effective number of neutrino species is however assumed 
to be ${\cal N}_{eff}=3.04$ through the whole suite. 
 
In this work we shall show how the Coyote emulator can be 
extended to models with a CPL DE state parameter [eq.~(\ref{w0wa})] 
without any significant accuracy degrade. To do so, we shall follow 
the pattern outlined in some previous papers 
\cite{Casarini2009,Casarini2010a,Casarini2010b} extending the approach 
of \cite{Francis2007}, also adding a specific evaluation of the shifts 
on the admitted range of $\sigma_8$, which depends on the $w_0$--$w_a$ 
values considered. Thanks to the width of the intervals explored by 
the Coyote simulation suite, the evaluated $\sigma_8$ limits (we shall 
plot them accurately in a forthcoming Section, for a specific choice 
of $\omega$'s and $n_s$) will be shown to include all physically 
relevant $w_0$--$w_a$ models and, in particular, those inside the 
2--$\sigma$ ellypse at the center of Fig.~\ref{fig:planckfig}. In 
connection with this work, we also provide a code, to be associated to 
the Coyote emulator, allowing one to predict spectra for $w_0$--$w_a$ 
cosmologies without significant accuracy degrade. 
 
In this work we shall also consider a fully different approach to  
detect fluctuation spectra, based on the halo model and first  
followed by \cite{Peacock1996}. Their analysis gradually evolved  
into the Halofit expressions \cite{Smith2003} furtherly extended by  
\cite{Takahashi2012}
and were also included in the CAMB code \cite{Lewis2000}. Along the
same pattern, massive neutrinos models, and an update for $f(R)$
modified gravity models \cite {Buchdahl1970,HuSawicki2007} were
included respectively in \cite {Bird2012} and \cite{Zhao2014}.
 
Beside Halofit, a standard halo model \cite  
{PeacockSmith2000,Seljak2000} fitting formula was provided by \cite  
{Mead2015} were baryons effects were tentatively included by using  
the simulations of \cite{VanDaalen2011}. Then \cite{Mead2016} went  
still forward, providing predictions for a number of more complex  
cases.  They include models with massive neutrinos, $f(R)$ gravity  
models, including those yielding Chamaleon or Vainshtein screening  
\cite {Vainshtein1972,DGP2000}, and DM--$\Phi$ coupling described by  
the equations  
\begin{equation}  
{T^{(c)}}^i_{j,i} = CT^{(c)} \Phi,j~,~~~ {T^{(d)}}^i_{j,i} = -  
CT^{(c)} \Phi,j~, \label{coupling}  
\end{equation}  
fulfilled by stress--energy tensors of DM and DE; apices $^{(c)}$  
and $^{(d)}$, respectively; these models were first simulated by  
\cite{maccio} and later by \cite{baldi} who also treated the case of~ 
$C(\Phi)$. In spite of such a wide range of cosmologies, the halo  
model exhibits its effectiveness by providing fair predictions, with  
an accuracy $\cal O$$(5-10\, \%)$. Although still far from the  
desired 1$\, \%$ limit, this is a significant step forward to cover  
the domain that experiments will be testing. 
 
Letting however apart these extensions, in association to this work we 
also provide a code to be used together with the CAMB package, 
enabling one to extend \cite{Takahashi2012} results for constant $w$, 
to any $w_0$--$w_a$ cosmology, without any appreciable accuracy 
degrade. 
 
The paper is organized as follows. In Section 2 we will review the 
mapping rule between $w_0-w_a$ model parameters and $w_{eq}$ \cite 
{Casarini2009}, providing explicit codes to implement it. In 
Section 3 we test the validity of the spectral equivalence using 
results from a set of N-body simulations. In Section 4 we discuss the 
physical origin of the spectral equivalence and compare results from 
different model mapping criteria. Section 5 is then devoted to a 
precise determination of CPL models to which the Coyote emulator can 
be extended.

\section{Non-Linear Matter Power Spectrum and Model Equivalence} 
Numerical studies of structure formation show signatures of DE state 
equation on a number of non--linear observables, as the halo mass 
function \cite{Bode2001, Klypin2003,Courtin2011}, the density profile 
of DM halos \cite {Klypin2003,Dolag2004,Deboni2013} and, namely, 
fluctuation spectra \cite{Ma1999,McDonald2006,Ma2007, Casarini2009, 
  Alimi2010,Jenning2010}. This comes as no surprise as DE alters the 
late linear growth. In turn, in an expanding background, the equations 
of motion for a system of collisionless particles are almost model 
independent if the linear growth factor is used as {\it time-variable} 
\cite{Nusser1998}. 
 
These arguments provide the physical basis to estimate the  
non-linear spectra of a given {\it target model} $\cal M$ at a given  
redshift $z$, from N-body simulations of an equivalent or {\it  
auxiliary model} ${\cal M}_{eq}$. In fact, \cite{Francis2007} showed  
that $w_0$--$w_a$ and constant--$w$ cosmologies exhibit close  
spectra at $z=0$ if, besides of sharing the cosmological parameters  
$\Omega_{b,m,r},~h, ~n_s$ and $\sigma_8$ (density parameters for  
baryons, matter, DM, reduced Hubble parameter, primeval scalar  
spectral index and m.s. fluctuation amplitude on the scale of 8$\,  
h^{-1}$Mpc at $z=0$, respectively), they also exhibit an identical  
comoving distance $d_0$ between $z=0$ and the last scattering  
surface (LSB). Moreover, according to \cite{Francis2007}, for some  
$w_0$--$w_a$ choices, there exists a greater redshift $\bar z$ where  
$\cal M$ and ${\cal M}_{eq}$ are again {\it equidistant} from the  
LSB. Spectral discrepancies, keeping within $\sim 1\, \%$ up to $k  
\simeq 3\, h\, $Mpc$^{-1}$ at $z=0$, grow up to $\cal O$$(5\, \%)$  
in the 0--$\bar z$ interval; they tend again to decrease in the  
proximity of $\bar z$, becoming wider and wider at still greater  
redshift. Aiming at 1$\, \%$ accuracy, \cite{Casarini2009} tried to  
extend the $z=0$ criterion at each $z > 0$, seeking $z$--dependent  
auxiliary models ${\cal M}_{eq}(z)$.  
 
First of all we require that {\it all ${\cal M}_{eq}(z)$ models share 
  target model $\Omega_{b,m,r}$ and $h$ parameter values at 
  $z=0$}. This soon implies ${\cal M}_{eq}(z)$ and $\cal M$ to share 
the reduced density parameters $\omega_{m,b,c} = \Omega_{m,b,c} h^2$  
at any $z$, even though $ \Omega_{m,b,c}(z)$ and $ H(z)$ are  
different. Then, if we want to determine the value of $w_{eq}$ and  
the present normalization amplitude of the power spectrum  
$\sigma_{8,eq}$ in model ${\cal M}_{eq}$, such that both models have  
the same non-linear power spectra at redshift $z$ as shown in \cite{Casarini2009}  
this requires imposing two conditions. 
 
First, by  
demanding that ${\cal M}$ and ${\cal M}_{eq}$ have the same distance  
to the last scattering surface one can infer the value of $w_{eq}$  
by numerically solving the equation:  
\begin{equation} 
  \int_z^{z_{lss}} {dz' \over E_{{\cal 
        M}_{eq}}(z')}=\int_z^{z_{lss}}{dz' \over E_{{\cal M}}(z')}~~~~~~~~ 
      {\rm with}~~ E^2_{\cal 
        M}(z)=\Omega_m(z)+\Omega_r(z)+\Omega_{d}(z)~. 
\label{firstcond} 
\end{equation} 
where $z_{lss}$ is the redshfit when last scattering surface occurs,  
$\Omega_m(z)=\Omega_m(1+z)^3$, $\Omega_r(z)=\Omega_r(1+z)^4$ and 
\begin{equation} 
  \Omega_{d}(z)= (1-\Omega_m-\Omega_r) \times ~\bigg\{~ 
  \begin{matrix} (1+z)^{3(1+w_{eq})} \hfill & \quad {\rm for} & {\cal M}_{eq} \cr 
    {\bf } \cr 
    (1+z)^{3(1+w_0+w_a)}\exp{[-3w_a z/(1+z)]} & \quad {\rm for} 
    & {\cal M}\end{matrix} 
\end{equation} 
 
Secondly, having determined $w_{eq}$ we can derive the value of  
$\sigma_{8,eq}$ in ${\cal M}_{eq}$ from the value of $\sigma_8$ in  
${\cal M}$ by imposing that the amplitude of the density  
fluctuations at the scale of $8$ Mpc h$^{-1}$ at redshift $z$ is the  
same in both models:  
\begin{equation}\label{secondcond} 
  \sigma_{8,eq}\frac{D_{{\cal M}_{eq}}(z)}{D_{{\cal  
  M}_{eq}}(0)}=\sigma_8\frac{D_{{\cal M}}(z)}{D_{{\cal M}}(0)}, 
\end{equation} 
where the linear growth factor is the growing mode solution to the  
linear perturbation equation: 
\begin{equation} 
 \frac{d^2D_{{\cal M}}}{d (\ln a)^2}+\frac{1}{2}\left[1-{3 w_{\cal 
       M}(a) \Omega_{d}(a)-\Omega_r(a) \over E^2_{\cal 
       M}(a)}\right]\frac{d D_{{\cal M}}}{d\ln 
     a}-\frac{3}{2}{\Omega_m(a) \over E^2_{\cal M}(a)}D_{{\cal M}}=0, 
\end{equation} 
 
For a given set of values of $w_0$ and $w_a$ these two conditions  
give at each redshift a unique value of $w_{eq}$ and $\sigma_{8,eq}$.  
 
\begin{figure}[th] 
\centering  
\vskip -.5truecm 
\includegraphics[width=.34\textwidth]{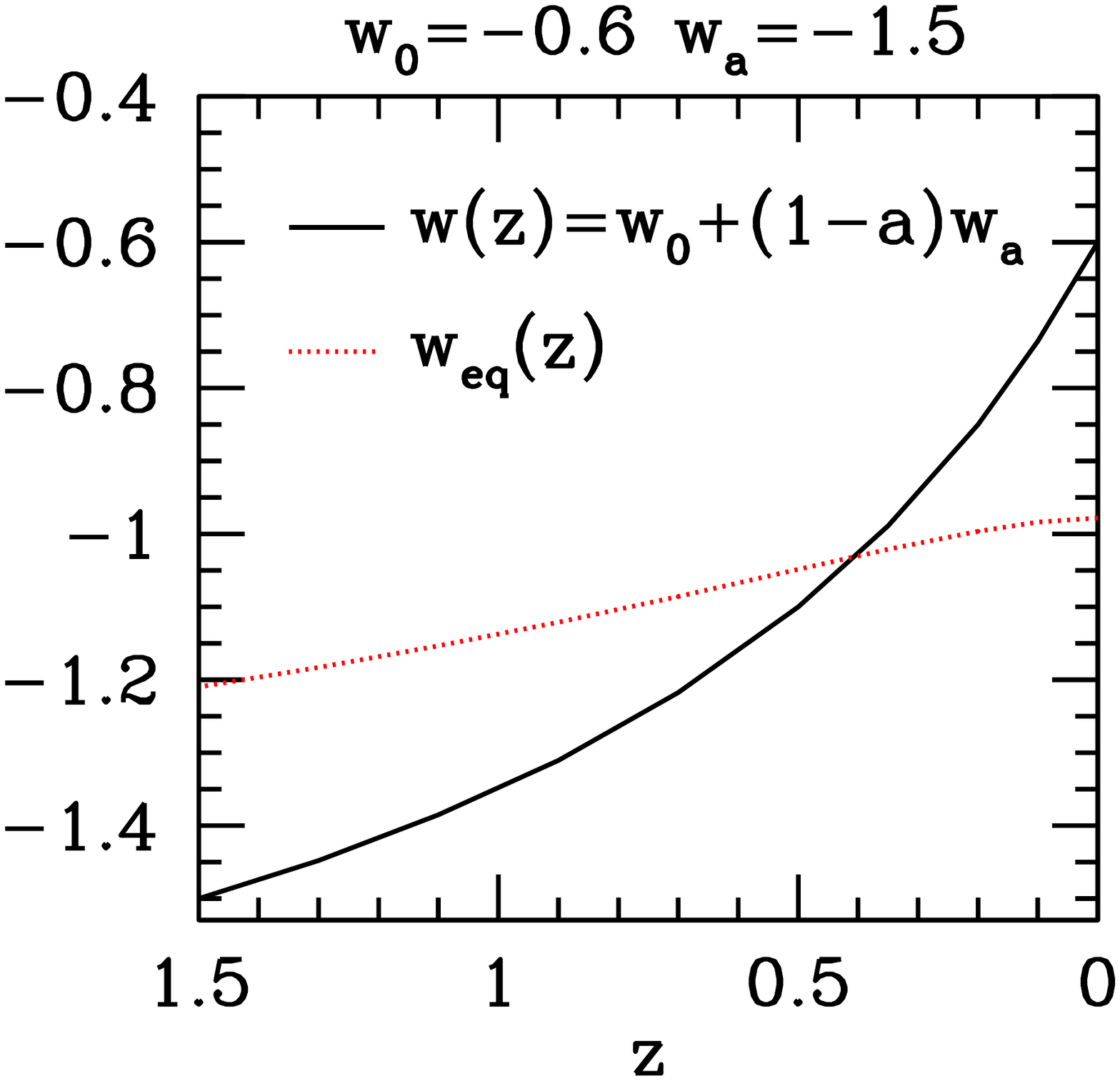} 
\includegraphics[width=.34\textwidth]{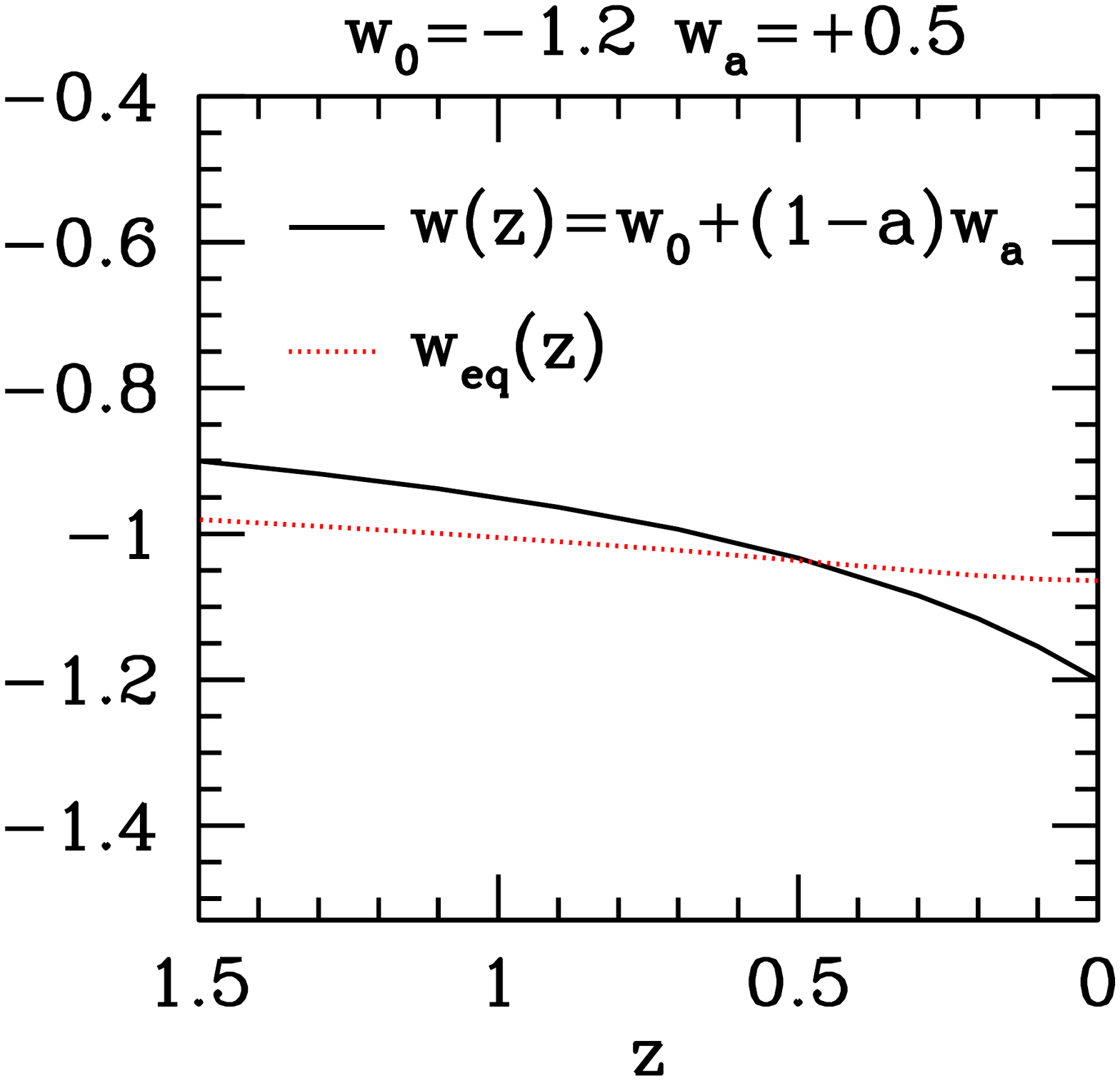} 
\vskip -2.2truecm 
\includegraphics[width=.342\textwidth]{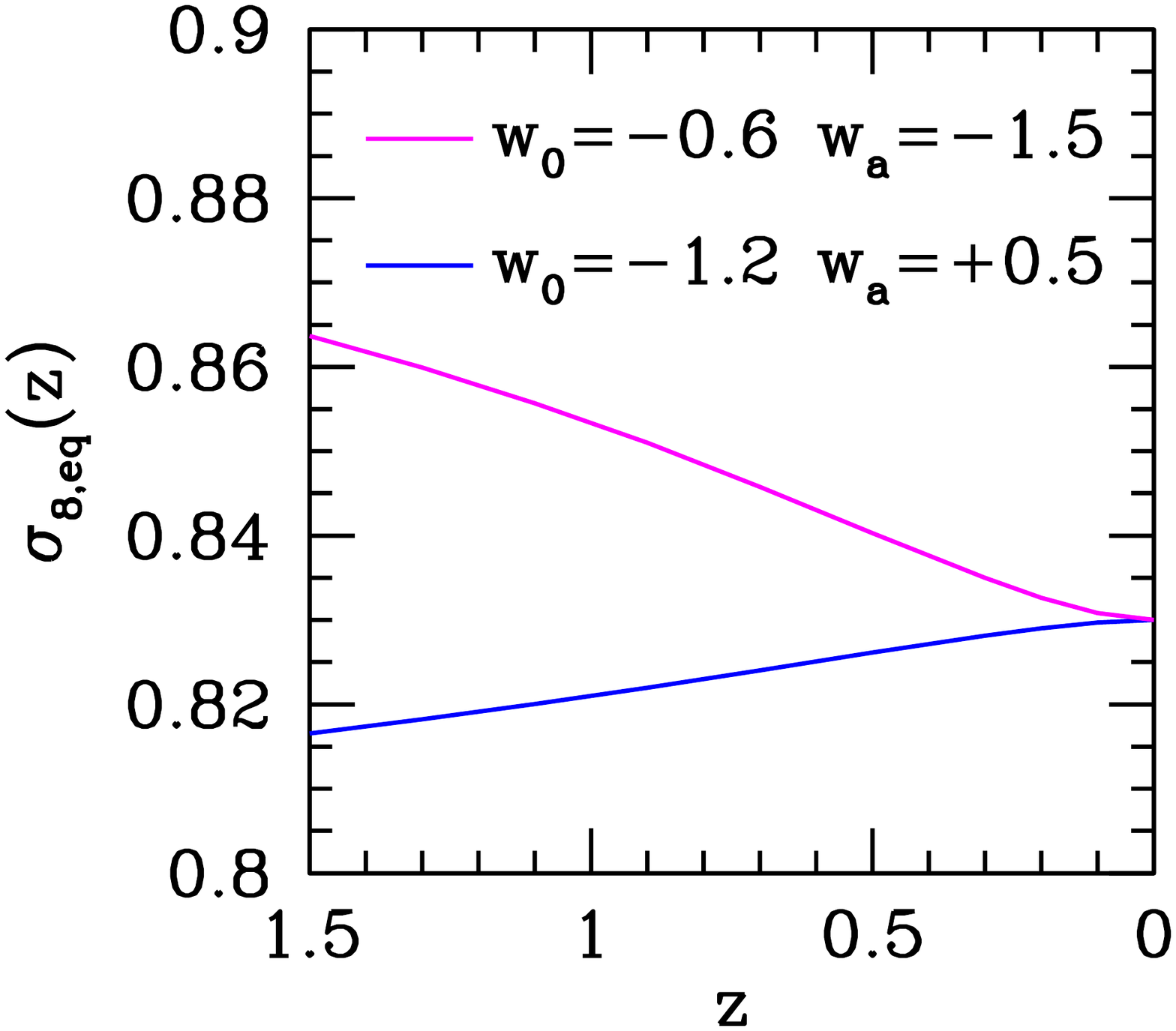} 
\vskip -2truecm 
 \caption{\label{fig:wew0} Auxiliary model $w_{eq}$ (top panels) and 
   $\sigma_{8,eq}$ (bottom panel) parameters, for two target DE models 
   with $w_0=-0.6$, $w_a=-1.5$ and $w_0=-1.2$, $w_a=0.5$ (solid lines 
   in top panels) respectively. Here we took $h=0.67$, $\Omega_m=0.32$ 
   and $T_0=2.725$ K.  } 
\end{figure} 
In former work \cite{Casarini2009,Casarini2010a,Casarini2010b} the 
validity of this criterion was tested for a wide range of models.  
Here, for the sake of example, Fig.~\ref{fig:wew0} shows tests 
concerning two different target models ${\cal M}$, close to the edge 
of $2\sigma$ contours of the smallest ellipse at the center of 
Fig.~\ref{fig:planckfig}. 
  
A {\sc fortran} program, called PKequal, computing the values of 
$w_{eq}$ and $\sigma_{8,eq}$ at a given redshift for any given target 
model ${\cal M}$ can be downloaded from the following link 
\url{https://github.com/luciano-casarini/pkequal}. This program 
was used, e.g., to obtain Fig.~\ref{fig:wew0}. 
 
\section{N-body Simulation Tests} 
In this Section we provide details on the tests run for this work, 
aimed at comparing the power spectra from N-body simulations of target 
and auxiliary models in and about the parameter range covered by the 
Coyote emulator. Particle distributions generated at $z=40$ by {\sc 
  graphic$–-$2} \cite{Bertschinger2001} were evolved down to $z=0$ by 
using the tree--code {\sc pdkgrav} \cite {Stadel2001}. Matter power 
spectra were then obtained at various redshifts through the code 
PMpowerM, included in the PM package described in 
\cite{KlypinHoltzman1997}, performing a Cloud-in-Cell interpolation of 
the particle distribution to reconstruct the density field on a 
uniform cartesian grid, with a maximum number of $N_g=2048$ cells. 
 
\begin{figure}[th] 
\centering  
\includegraphics[width=.85\textwidth]{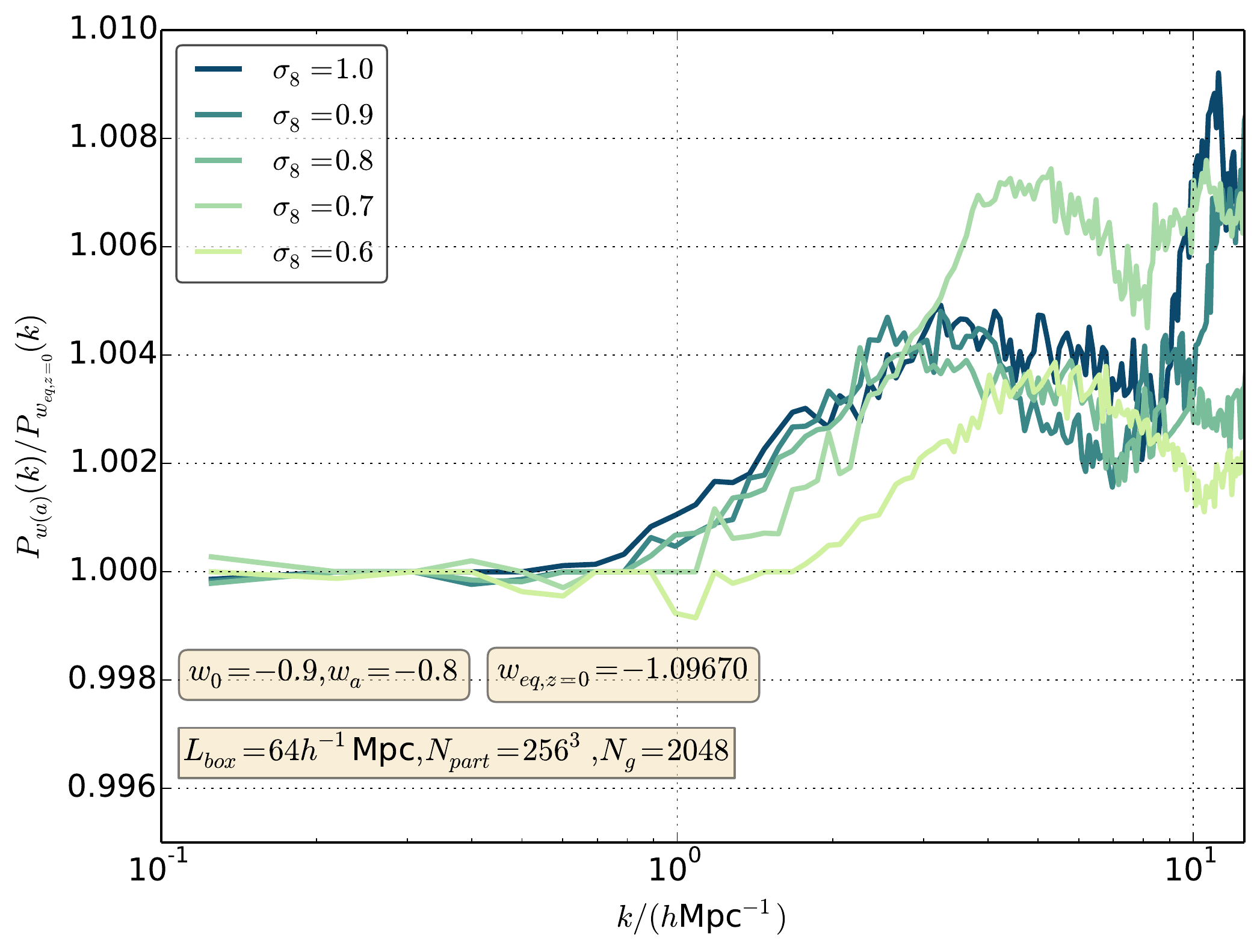} 
\includegraphics[width=.85\textwidth]{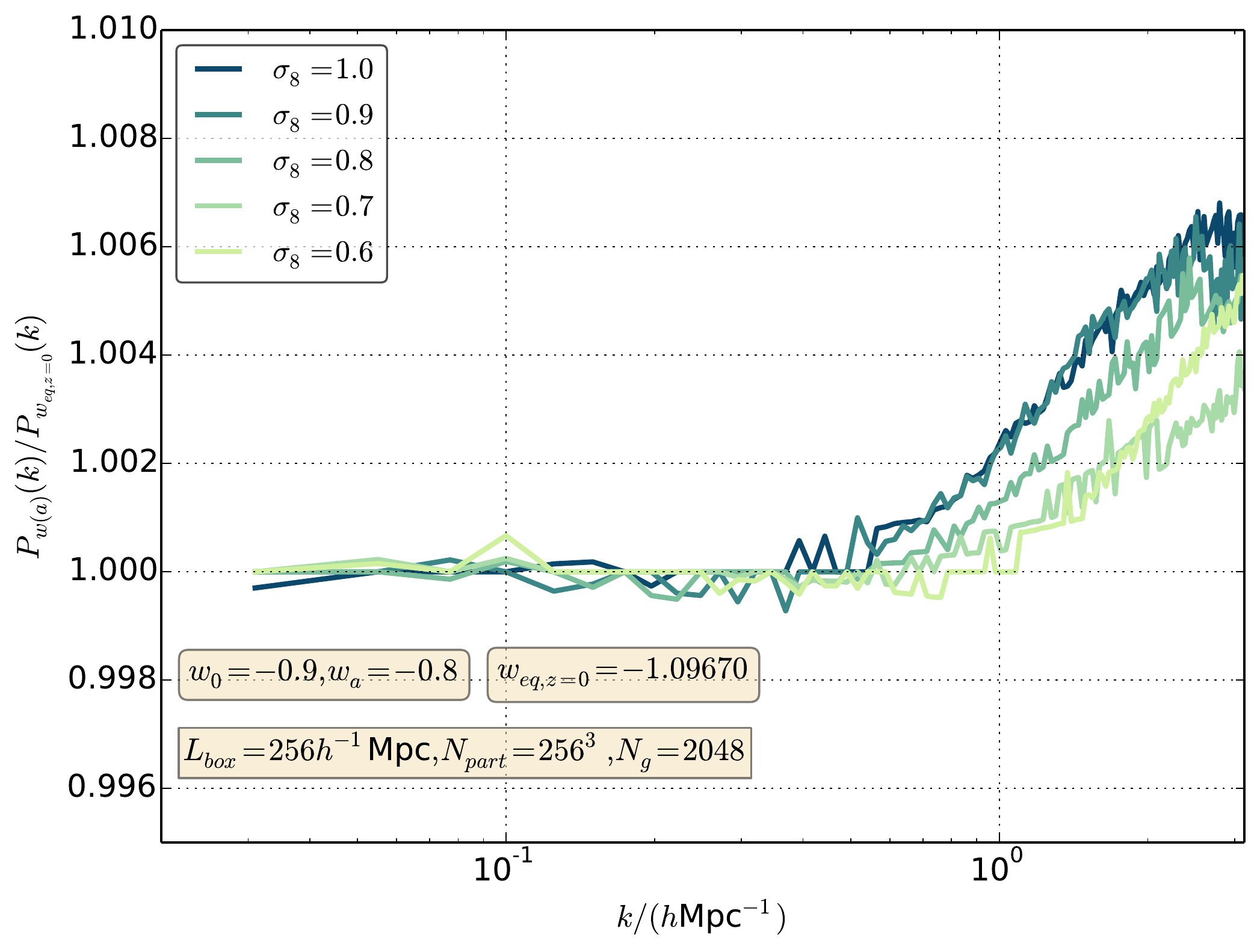} 
\caption{\label{fig:rats8} Ratio of the non-linear matter power 
  spectra at $z=0$ of target and auxiliary models from simulations 
  with $64\,h^{-1}$ Mpc (top panel) and $256\,h^{-1}$ Mpc (bottom 
  panel) box size respectively for different value of $\sigma_8$.} 
\end{figure}

\begin{figure}[th] 
\centering  
\includegraphics[width=.85\textwidth]{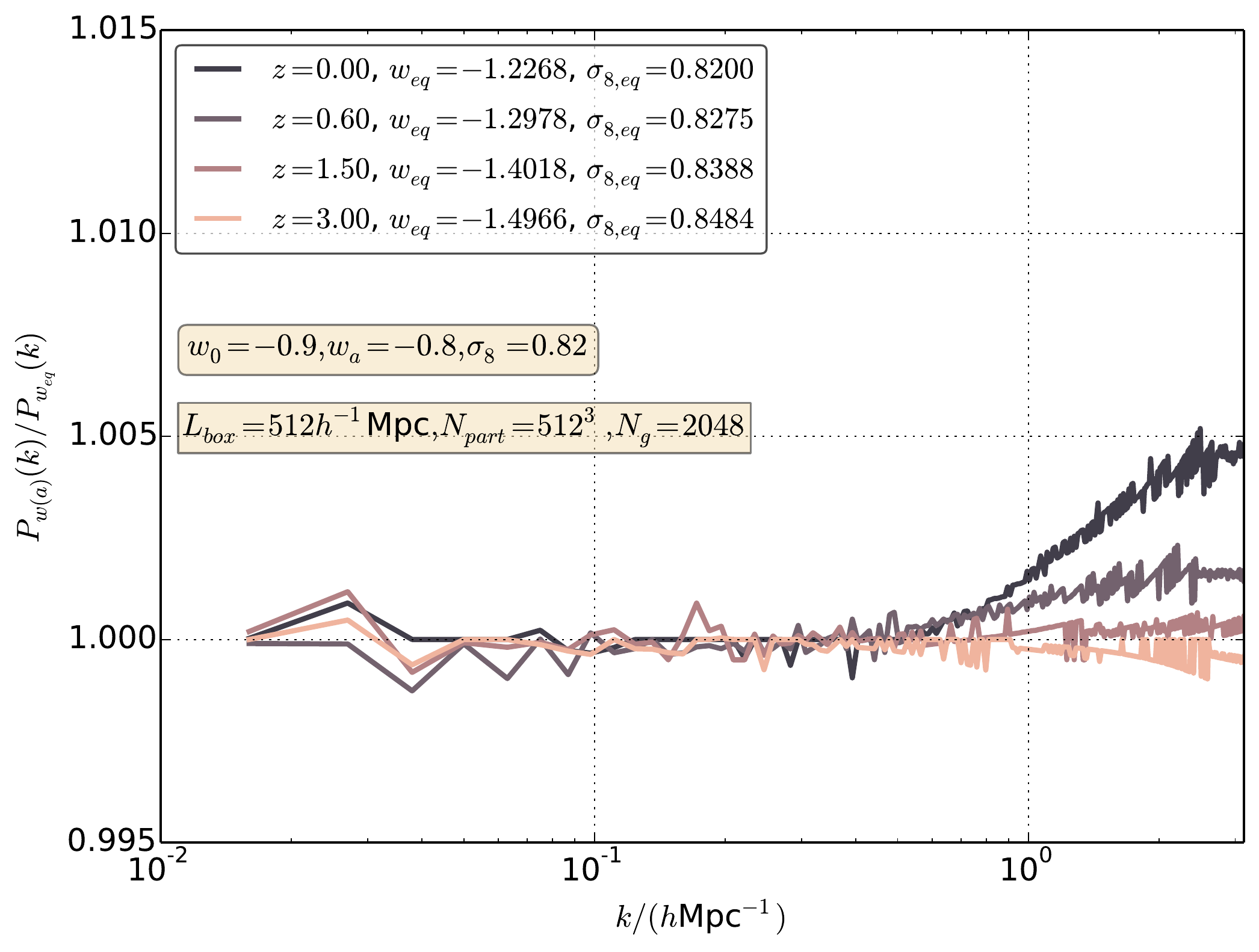} 
\caption{\label{fig:ratz0} Ratio of the non-linear matter power 
  spectra of target and auxiliary models from simulations with 
  $512\,h^{-1}$ Mpc box size at $z=0,0.6,1.5$ and $3$ respectively.} 
\end{figure}

Simulations were run in boxes with various box sizes and particle 
numbers, yielding different values for the Nyquist frequency $k_{\rm 
  Ny}\equiv \pi N_p^{1/3}/L$, mostly considered a fair estimate of the 
maximum $k$ where simulations are numerically reliable. At an accuracy 
level $\ll 1\, \%$, this however needs to be verified. The simulations 
used to obtain the results in Figures \ref{fig:rats8} and 
\ref{fig:ratz0} are shown in the Table 
$$ 
\begin{matrix} ~~~ & L/h^{-1}{\rm Mpc}& ~~~ & N_p & ~~~ & k_{Ny}/h{\rm 
    Mpc}^{-1} \cr {\rm } \cr S(64,256)\hfill &~~~ 64 & ~~~ & 256^3 & ~~~ & 
  12.5 \cr S(256,256)\hfill &~~~ 256 & ~~~ & 256^3 & ~~~ & 3.1 \cr 
  S(512,512)\hfill &~~~ 512 & ~~~ & 512^3 & ~~~ & 3.1 \cr  
\end{matrix} 
$$ 
Their names shown in the first column are meant to remind $L$ and $N_p$. 
 
This selection of simulation parameters is aimed to test the effects 
of longer wavelength modes and small scale resolution on the 
discrepancies between target and auxiliary models, for which we aim at 
reaching a precise scale and redshift dependence. 
 
 
Here below we shall illustrate results for a CPL target model with 
$\Omega_b=0.048$, $\Omega_m=0.7$, $h=0.70$, $n_s=0.966$ and, namely, 
$w_0=-0.9$ and $w_a=-0.8$, treated also in Figure \ref{fig:wew0}. 
Those related to the other extreme of the likelihood ellipse are 
mostly analogous and we shall just outline the differences we found. 
 
In Coyote's range $0 \leq z \leq 1$ the auxiliary models of such 
cosmology have state parameter ranging from $w_{eq}=-1.0967$ ($z=0$) 
to -1.1863 ($z=1$), as also shown in Fig.~\ref{fig:wew0}. These values 
are close to the center of the $w$-interval covered by the Coyote 
emulator. A set of $\sigma_8$ values, ranging from 0.6 to 1, were 
considered. Results will be shown in Figure \ref{fig:rats8}. 
 
We however went forward and, in Figure \ref{fig:ratz0}, will show 
results for redshifts up to $z=3\, ,$ although for a unique target 
model with $\sigma_8 = 0.82\, $. 
 
From all simulations we derive ratios between matter power spectra of 
target and auxiliary models at $z=0$ where, according to previous 
analysis, discrepancies reach their maximum.  In the top panel of 
Figure~\ref{fig:rats8} we plot such ratio, as derived from 
$S(64,256)$, up to $k_{\rm Ny}$ at $z=0$ for different values of 
$\sigma_8$\footnote{Notice that at $z=0$ we have 
  $\sigma_{8}=\sigma_{8,eq}$}. Up to $k = 3\, h\, $Mpc$ ^{-1}$, for 
$0.8 \leq \sigma_8 \leq 1$, discrepancies essentially overlap. For 
$\sigma_8 = 0.7$ (0.6) they are slightly (significantly) smaller. 
Until $k \lesssim 1.5\, \, h$Mpc$^{-1}$ they never exceed $0.3\, \%$; 
they then approach $\simeq 0.4\, \%$ ($\simeq 0.5\, \%$) at $k \sim 
2.2\, h$ Mpc $^{-1}$ ($3\, h$ Mpc $^{-1}$). It should be also outlined 
that, at $k > 3\, h$ Mpc $^{-1}$, the discrepancy growth tends to 
stop, never bypassing $1\, \%$. 
 
The bottom panel of Fig.~3 is derived from $S(256,256)$. Longer 
wavelength modes are therefore included, at the expenses of a smaller 
resolution, although $k_{Ny}$ however exceeds $3\, h\, $Mpc$^{-1}$. 
Model discrepancies appear slightly greater here than in $S(64,256)$, 
approaching $0.7\, \%$ at $k = 3\, h\, $Mpc$ ^{-1}$ while however 
keeping below $0.5\, \%$ for $k < 2\, h\, $Mpc$ ^{-1}$.  This is 
associated to a start of discrepancies at a slightly lower $k$ than in 
in $S(64,256)$. 
 
There seem to be scarce doubts that the latter tiny effect arises from 
the contribution of long wavelength discrepancies. On the contrary, 
comparison with the next Figure will confirm that the bulk of the 
former slight increase is to be attributed to $k_{Ny}$ being already 
too close.  Even including such a tiny numerical noise effect, 
however, the $0.5\, \%$ limit for $k < 2\, h\, $Mpc$ ^{-1}$ is kept. 
 
Let us finally comment the discrepancies in $S(512,512)$. Although 
restricted to $\sigma_8 = 0.82$ at $z=0$, Figure \ref{fig:ratz0} 
shows, first of all, how discrepancies decrease with increasing $z$, 
becoming apparent only for $z \lesssim 1.5$. It is however worth 
outlining that, in the case of the model with $w_0=-1.2$, $w_a=0.5$, 
discrepancies do not dye so early. This is due to the greater values 
of $w_{eq}(z)$, only partially balanced by $\sigma_{8,eq}$ decreasing 
--instead of increasing, see Figure 2-- with $z$. Even in this case, 
however, no residual discrepancy can be appreciated above $z \simeq 
2\, $. 
 
Apart of that, we may compare the discrepancies at $z=0$ with those 
for $\sigma_8 = 0.8$ in the previous Figure. This allows us to 
appreciate, first of all, that doubling the volume side does not 
modify at all the $k$ range above which discrepancy exhibit their mild 
increase, i.e, longer wave have no more impact on that. Then, at $k 
\sim 2.5$--$3 \, h\, $Mpc$^{-1}$, we recognize almost identical 
details, hence presumably deriving from the approaching of identical 
Nyquist frequencies. Although bearing no impact on the discrepancy 
limits for $k \leq 2 \, h\, $Mpc$^{-1}$, this makes us believe that, 
in this region, the top panel of Fig.~\ref{fig:rats8} may be more 
reliable than the bottom one. 
 
Altogether we can conclude that residual model discrepancies increase 
with $k$ (while decreasing with $z$). Being mostly interested here 
just to $k \leq 2\, h\, $Mpc$^{-1}$ (above such $k$ baryon physics 
requires corrections $>\sim 1\, \%$) model discrepancies were never 
seen to exceed $0.5\, \%$. However, if one wishes to use the procedure 
up to $k \sim 3\, h\, $Mpc$^{-1}$, a safe accuracy limit can be $1\, 
\%$. 
 
\section{Physical Origin of Spectral Equivalence} 
In this Section we shall compare other possible options with the 
equivalence criterion set by eq.~(\ref{firstcond}), which can be also 
restated as requiring that equal conformal times have elapsed from the 
LSS to the relevant redshift $z$.  
 
It has been known since long that the conformal time sets the best 
metric on the time coordinate, over scales involved in the cosmic 
expansion. Such metric is often used in linear codes (e.g. by CAMB) 
and primeval sonic waves are (quasi--)periodical in respect to 
it. Then, as mentioned in \cite{Nusser1998}, the linear growth factor 
defines a time variable that rescale the dynamical equations of N-body 
particles in a form that is nearly independent of the underlying 
cosmological model. Thus, once two models share the same linear growth 
histories, when N-body particles begin to form non--linearities, the 
similarity is preserved and eventually leads to similar power spectra.

To show this more clearly, we test the spectral equivalence for two 
tentative additional DE model mapping criteria, defined as the 
following.  
 
For a given set of values of $w_0$ and $w_a$ we infer $w_{eq}$ by 
requiring: 
\begin{itemize} 
\item equality of the linear growth factors at the redshift of 
  interest, 
\begin{equation}\label{dplus} 
D_{{\cal M}_{eq}}(a)=D_{{\cal M}}(a); 
\end{equation} 
\item equality of the integral of the linear growth factors up to the 
  redshift of interest, 
\begin{equation}\label{idplus} 
\int_0^a D_{{\cal M}_{eq}}(a')da'=\int_0^a D_{{\cal M}}(a')da'. 
\end{equation} 
\end{itemize} 
 
\begin{figure}[th] 
\centering 
\includegraphics[width=.7\textwidth]{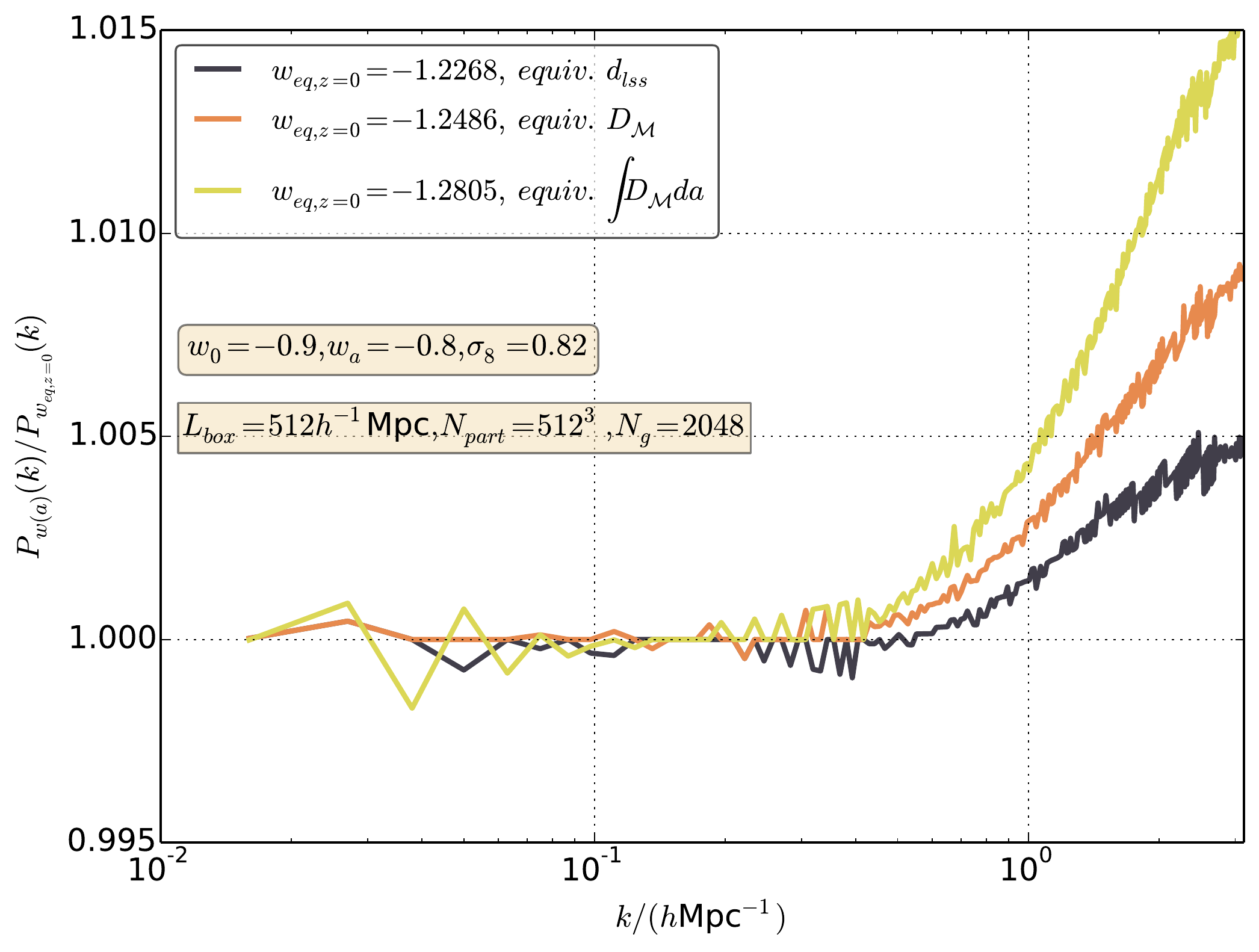} 
\includegraphics[width=.7\textwidth]{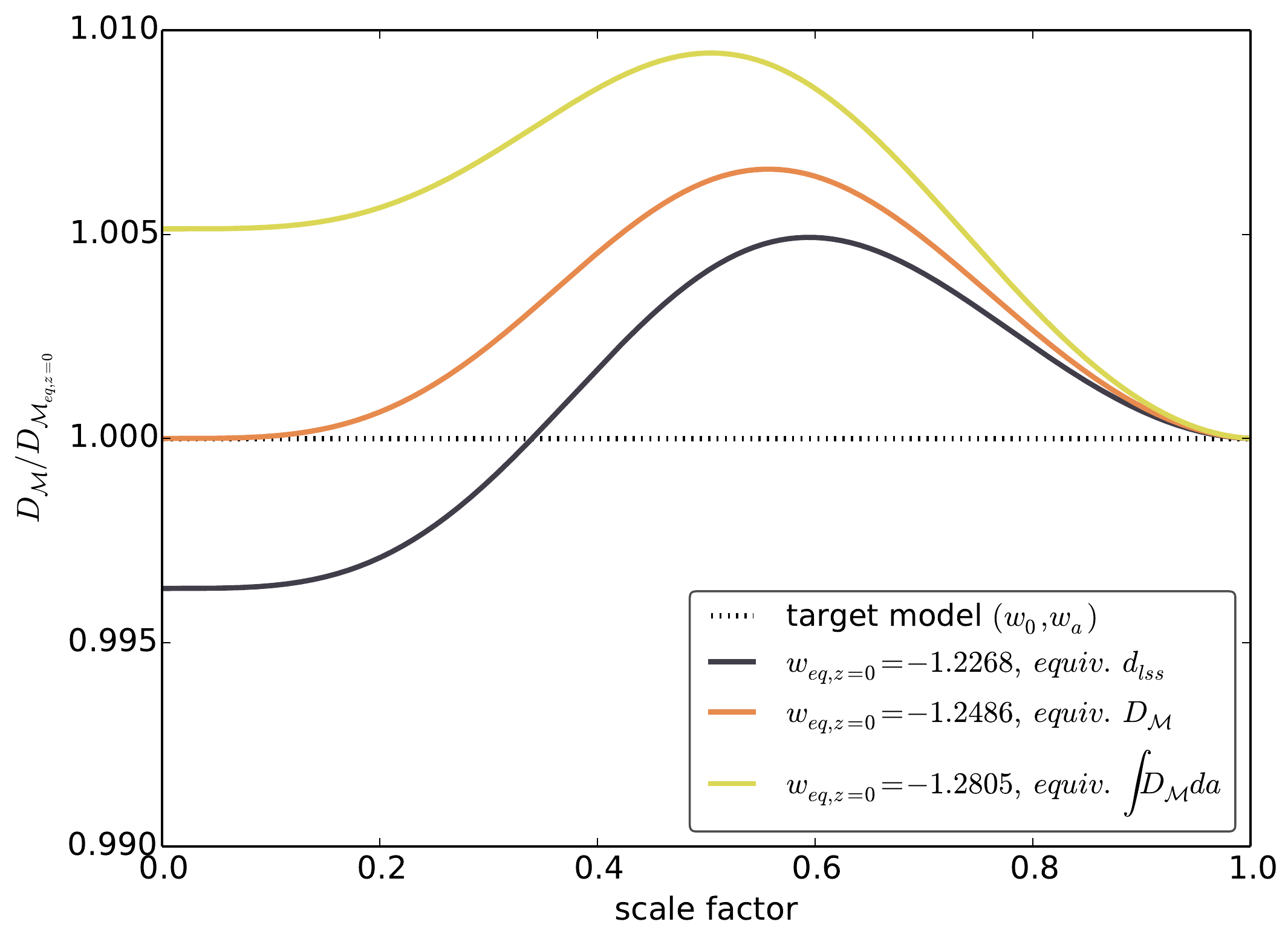} 
\caption{\label{fig:App1} Comparison between different equivalence criteria at $z=0$.} 
\end{figure} 
We consider again a target model with $w_0=-0.9$ and $w_a=-0.8$. At 
$z=0$ it is mapped into an auxiliary model with constant equation of 
state $w_{eq}=-1.2268$ [$-1.2486$, $-1.2805$] for the criterion 
(\ref{firstcond}) [(\ref{dplus}), (\ref{idplus})]. 
To test such two additional criteria, we run {\it ad--hoc} simulations 
for each {\it equivalent} model defined according to them. We used a 
$512\,h^{-1}$ Mpc box size, with $512^3$ particles, and set 
$\sigma_8=0.82$; the other cosmological parameters are kept equal. 
 
In Figure~\ref{fig:App1} we plot the ratio of non-linear power spectra 
at $z=0$ (top panel) of target and auxiliary models for the three 
different criteria and the ratio of the normalized linear growth 
factors (bottom panel). This shows that the differences between the 
non-linear spectra in the high-$k$ interval are correlated with 
deviations of the linear growth histories between target and auxiliary 
models. The mapping rule based on the equality of the distance to last 
scattering surface (conformal times) shows the smallest deviations 
between non-linear spectra as well as the smallest differences of the 
linear growth factors across the entire cosmic history. In contrast, 
the criterion based on the integral of the growth factors leads to the 
largest discrepancies of the spectra as well as the large deviations 
between the linear growth histories. This clearly indicates that 
whenever two cosmologies share similar growth factors they should also 
have similar non-linear power spectra. 
 
This criterion suggests a mapping rule that, in principle, could be 
extended beyond the case of DE parametrizations, e.g. to Modified 
Gravity models. This might deserve an accurate future study. 
 
A point outlined in the top panel of Figure \ref{fig:rats8} is that 
discrepancies between target and auxiliary models apparently stabilize 
even when we extend the spectral comparison up to large $k$'s. 
Admittedly, this part of computed spectra has no direct 
phenomenological meaning as, there, baryons cannot be neglected. An 
open question, however, is whether non--linear baryon dynamics 
necessarily leads to differentiating otherwise equivalent models 
which, e.g., share $\Omega_b$ and $n_s$ values; and, if it does so, 
which are the origin and the size of the expected discrepancies. 
 

\section{CPL parameter range of the Coyote emulator} 
 
\begin{figure}[th] 
\centering  
\vskip -3truecm 
\includegraphics[width=.7\textwidth]{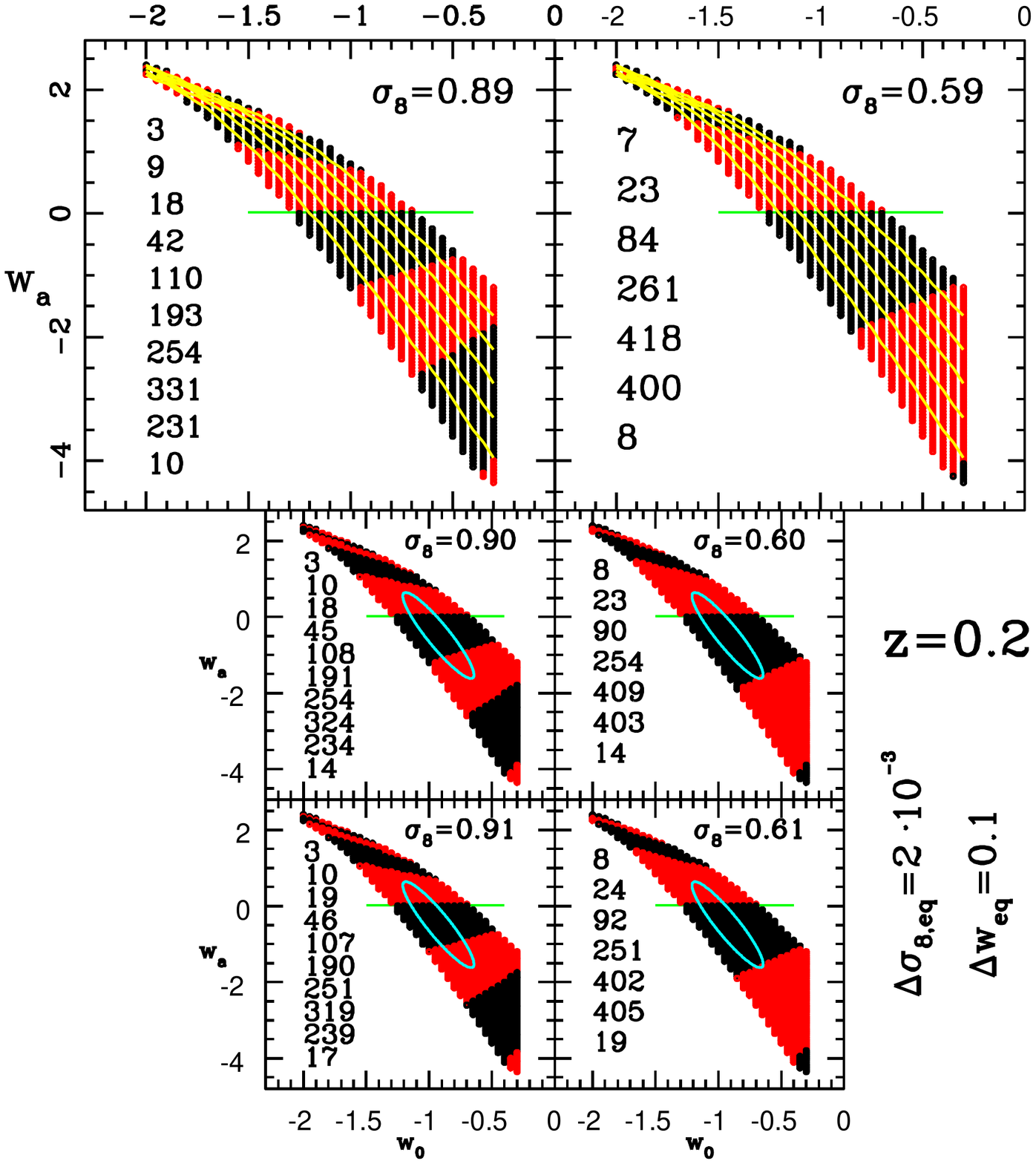} 
\vskip -4truecm 
\includegraphics[width=.7\textwidth]{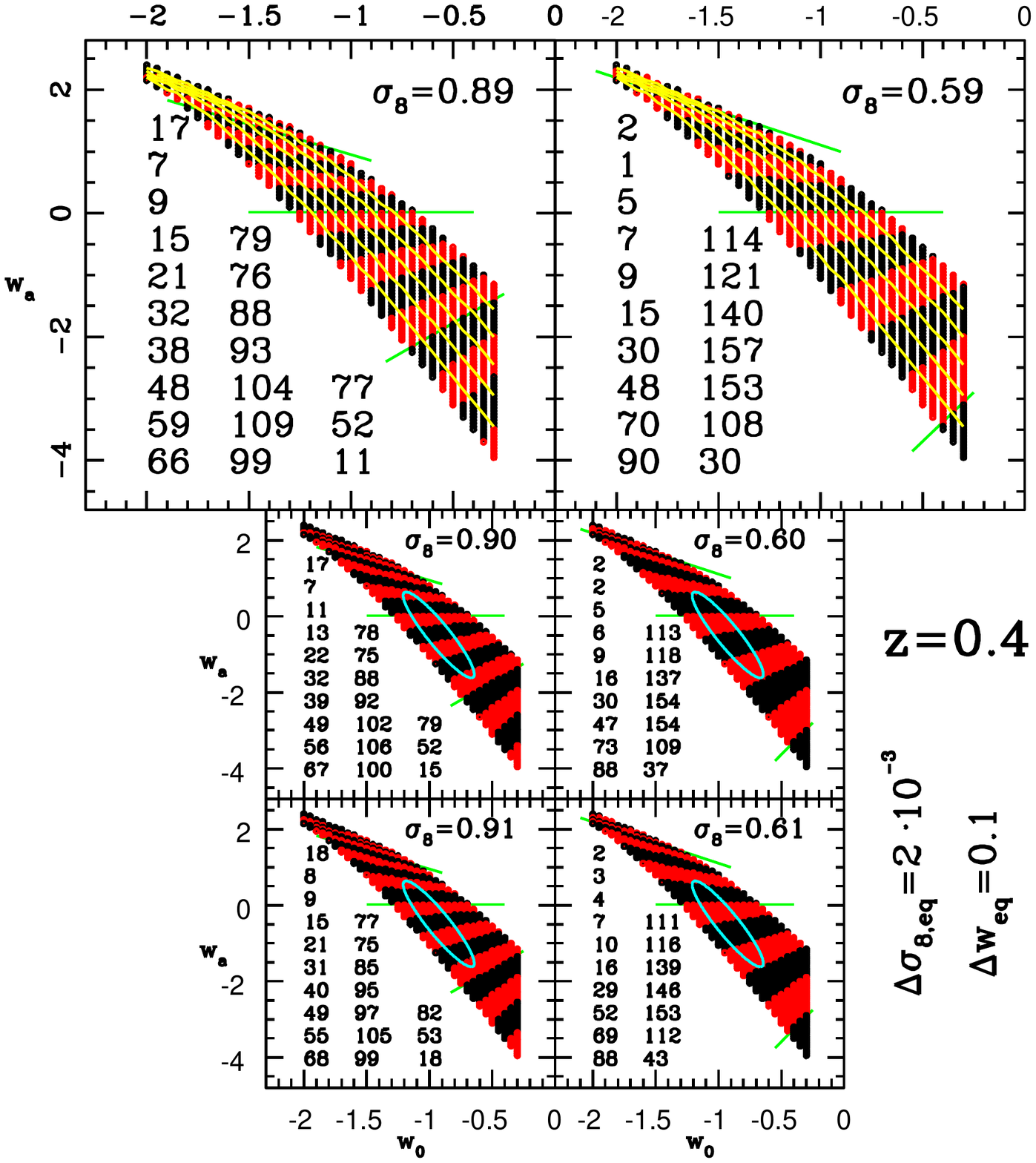} 
\vskip -3truecm 
\caption{\label{fig:equi} Region of the $w_0-w_a$ plane with auxiliary 
  models in the Coyote suite for $z=0.2$ (top panel) and $0.4$ (bottom 
  panel). The yellow lines in the larger plots give the values of 
  $w_{eq}$ varying in the range $-1.3<w_{eq}<-0.7$ (bottom to top) in 
  steps of $\Delta w_{eq}=0.1$. The horizontal green lines indicate 
  the case $w_a=0$ which corresponds to auxiliary models with 
  $w_{eq}=w_0$ and $\sigma_{8,eq}=\sigma_{8}$. The red and black areas 
  below (above) the horizontal green lines indicates increments 
  (decrements) in the value of $\sigma_{8,eq}$ of the auxiliary model 
  in steps of $\Delta\sigma_{8,eq}=0.002$. The additional green lines 
  in the case of $z=0.4$ mark the shifts in increments (decrements) in 
  steps of $\Delta\sigma_{8,eq}=0.01$.} 
\end{figure} 
 
\begin{figure}[th] 
\centering  
\vskip -3truecm 
\includegraphics[width=.7\textwidth]{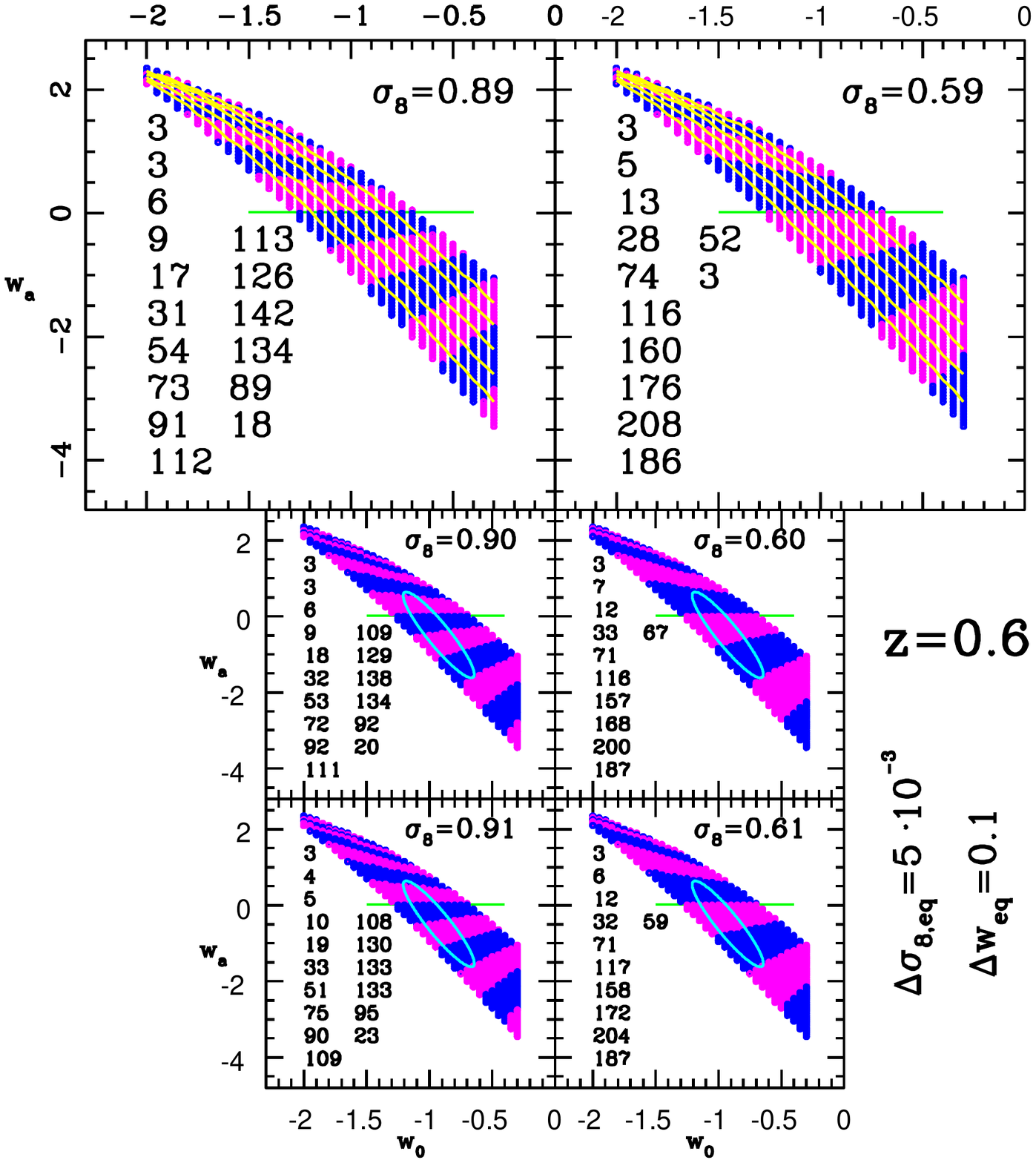} 
\vskip -4truecm 
\includegraphics[width=.7\textwidth]{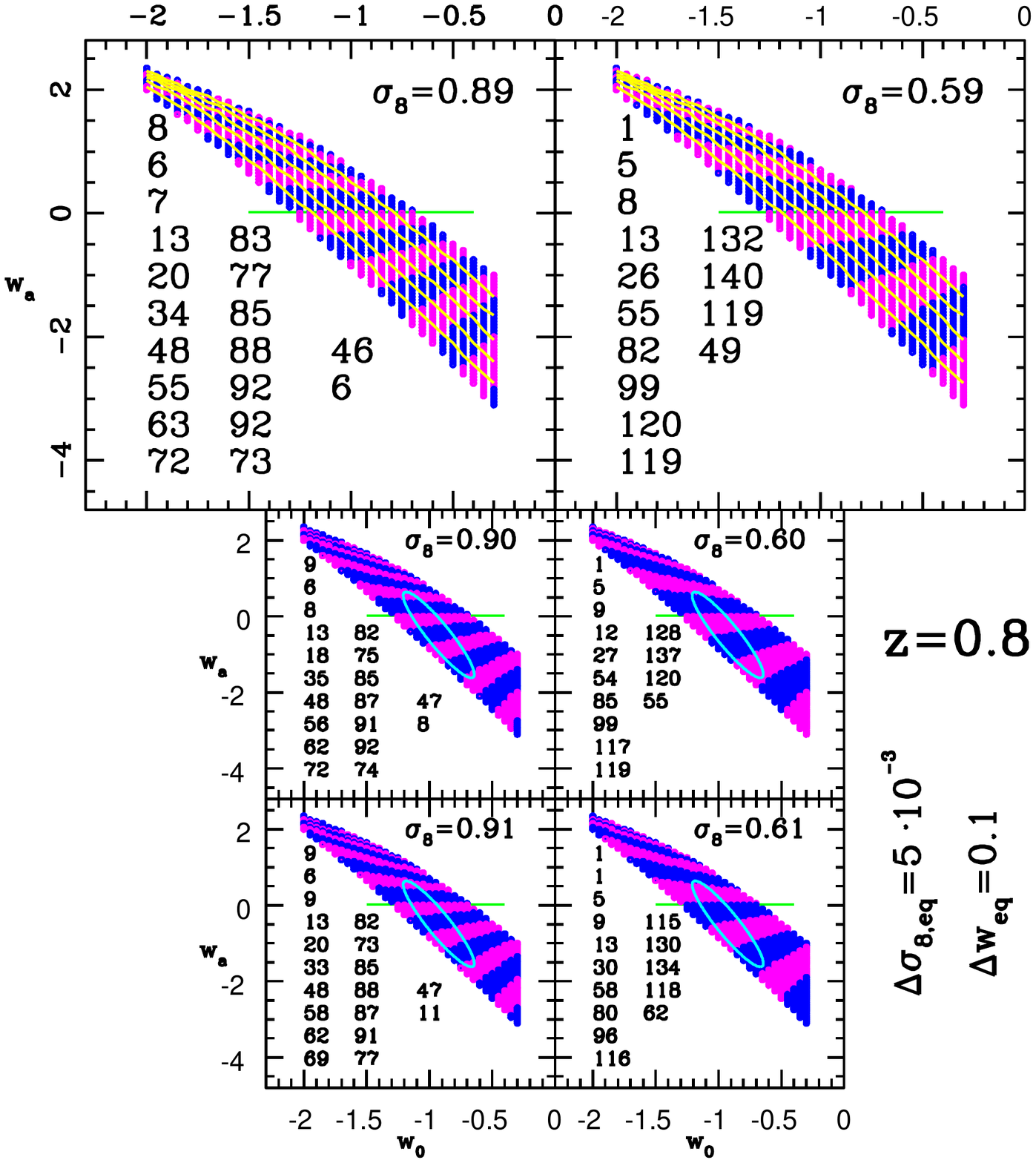} 
\vskip -3truecm 
\caption{\label{fig:equi1} As in Figure~\ref{fig:equi} for $z=0.6$ 
  (top panel) and $0.8$ (bottom panel) with increments (decrements) in 
  the value of $\sigma_{8,eq}$ of the auxiliary model in steps of 
  $\Delta\sigma_{8,eq}=0.005$.} 
\end{figure}

\begin{figure}[th] 
\centering  
\includegraphics[width=.7\textwidth]{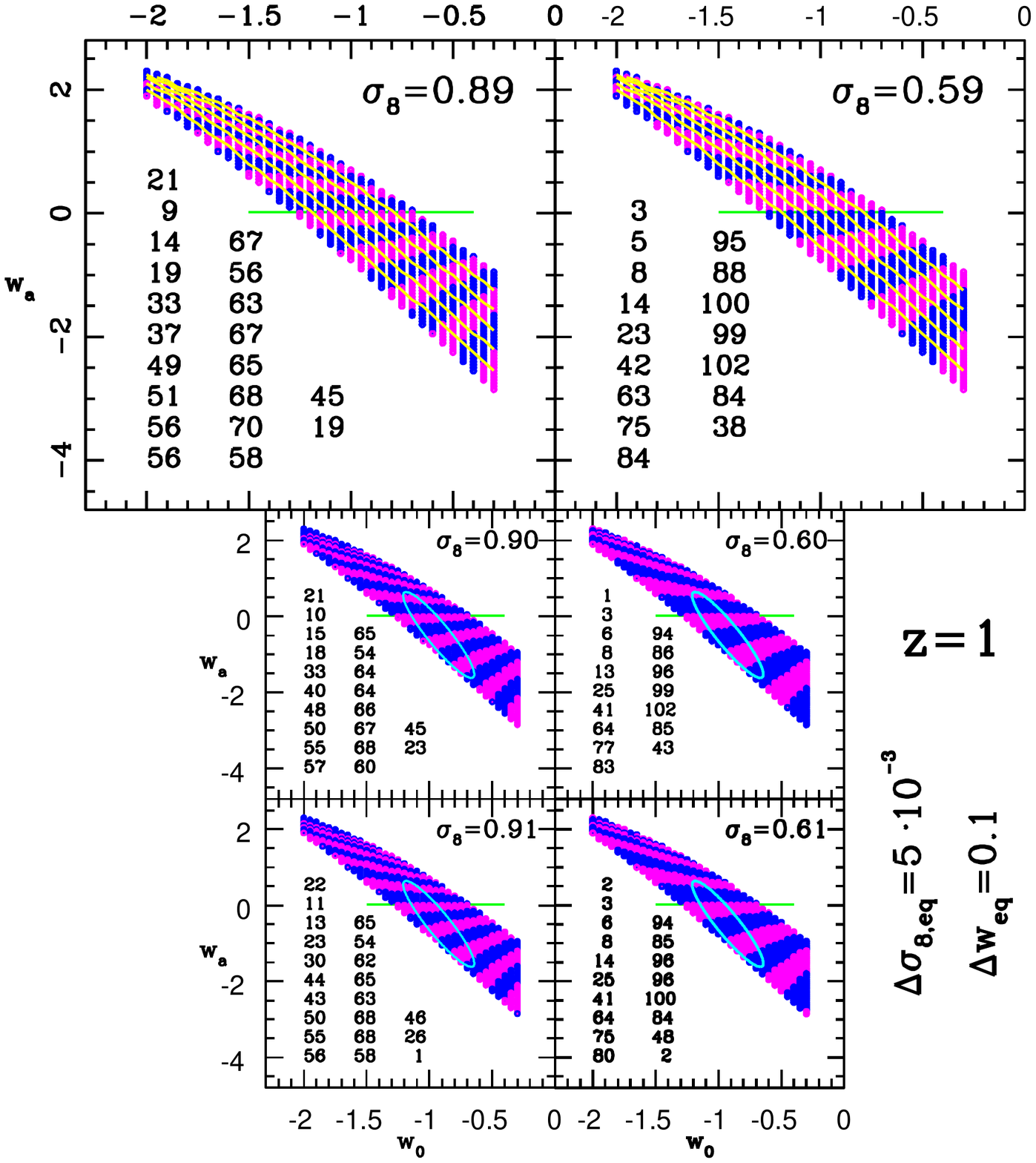} 
\vskip -3truecm 
\caption{\label{fig:equi2} As in Figure~\ref{fig:equi1} for $z=1$.} 
\end{figure} 
 
The Coyote emulator provides non-linear power spectra for DE models 
with constant equation of state. Thus, we can use the spectral model 
equivalence described in the previous Section to extend the 
predictions to the CPL parameter space. Here, we determine the range 
of values of $w_0$ and $w_a$ that for a given value of $\sigma_8$ and 
at a given redshift have an auxiliary model with values of $w_{eq}$ 
and $\sigma_{8,eq}$ in the range covered by the Coyote emulator. 
 
Hereafter, we set $h=0.70$, $\omega_b=0.048$, $\Omega_m=0.7$ and 
$n_s=0.966$ which are well within the range probed by the Coyote 
suite. 
 
At $z=0$, we find that all values of $w_0$ and $w_a$ within the 
$2\sigma$ countours of the combined Planck data analysis shown in 
Figure~\ref{fig:planckfig} have a counterpart in the Coyote suite. At 
higher redshift the availability of auxiliary models depends on 
redshift the specific values of $w_0$, $w_a$ and $\sigma_8$ as 
summarized in Figure~\ref{fig:equi} for $z=0.2$ (top panel) and $0.4$ 
(bottom panel), Figure~\ref{fig:equi1} for $z=0.6$ (top panel) and 
$0.8$ (bottom panel) and Figure~\ref{fig:equi2} for $z=1.0$. In each 
panel the various plots correspond to different $\sigma_8$. We 
specifically focus on values which are at the extreme end of the 
Coyote interval, since it is in these cases that target models may 
corresponds to auxiliary ones with $\sigma_{8,eq}$ values outside the 
Coyote range. Thus, in the figures we plot region of the CPL parameter 
space for $\sigma_8=0.59,0.60,0.61,0.89,0.90$ and $0.91$ respectively. 
 
The information encoded in these plots can be read as the 
following. Each panel show a region of the $w_0$-$w_a$ plane (roughly 
corresponding to the $2\sigma$ region constrained by the Planck data, 
see Figure~\ref{fig:planckfig}) that maps corresponding values of 
$w_{eq}$ probed by the Coyote emulator. In each panel the values of 
$w_{eq}$ can be read from the larger plots as a series of nearly 
parallel yellow lines varying in the range $-1.3<w_{eq}<-0.7$ in steps 
of $\Delta w_{eq}=0.1$ from bottom to top. The horizontal green lines 
correspond to the case $w_a=0$, where $w_{eq}=w_0$ and 
$\sigma_{8,eq}=\sigma_{8}$. The alternating black and red bands above 
(below) the horizontal green lines correspond to increments 
(decrements) of the $\sigma_{8,eq}$ value by 
$\Delta\sigma_{8,eq}=0.002$ for $z=0.2$ and $0.4$, and 
$\Delta\sigma_{8,eq}=0.005$ for $z=0.6,0.8$ and $1.0$ 
respectively. The smaller plots in each panel show the case 
corresponding to $\sigma_{8}$ values that are at the limit (or just 
outside) the range covered by the Coyote emulator. Here, only certain 
regions of the $w_0-w_a$ parameter space have auxiliary models in the 
Coyote suite, while other have no counterpart includes models that are 
within the $2\sigma$ contours from the combined Planck data analysis 
(represented in the figure by the elliptical contour). Nonetheless, it 
is worth noticing that such regions are associated with extreme values 
of $\sigma_{8}\sim 0.6$ and $0.9$ that are more than $2\sigma$ away 
from Planck best-fit value. 
 
The various plots have been realized through a discrete sampling of 
the CPL parameter space in steps of $0.05$ in $w_0$ and $w_a$. The 
list of numbers in each plot specifies the number of $w_0$-$w_a$ pairs 
for each band of $\sigma_{8,eq}$ values. In the case $z=0.4$ we have 
two additional green lines above and below the horizontal one which 
mark the shift in the increment (decrement) of $\sigma_{8,eq}$ by 
$\Delta\sigma_{8,eq}=0.01$. Notice that the limits shown in 
Figure~\ref{fig:equi} to ~\ref{fig:equi2} hold when the remaining 
cosmological parameters are set to fiducial values; we however tested 
that varying $\Omega_m$ and $h$ within current observational 
uncertainties induces negligible differences. We can conclude that, up 
to $z=1$, the bulk of physically significant DE models with CPL 
parametrization have a counterpart in the ensemble of models 
originally covered by the Coyote emulator. 
 
\section{Comparison with HMcode}  
 
\begin{figure}[th]  
\centering  
\includegraphics[width=.8\textwidth]{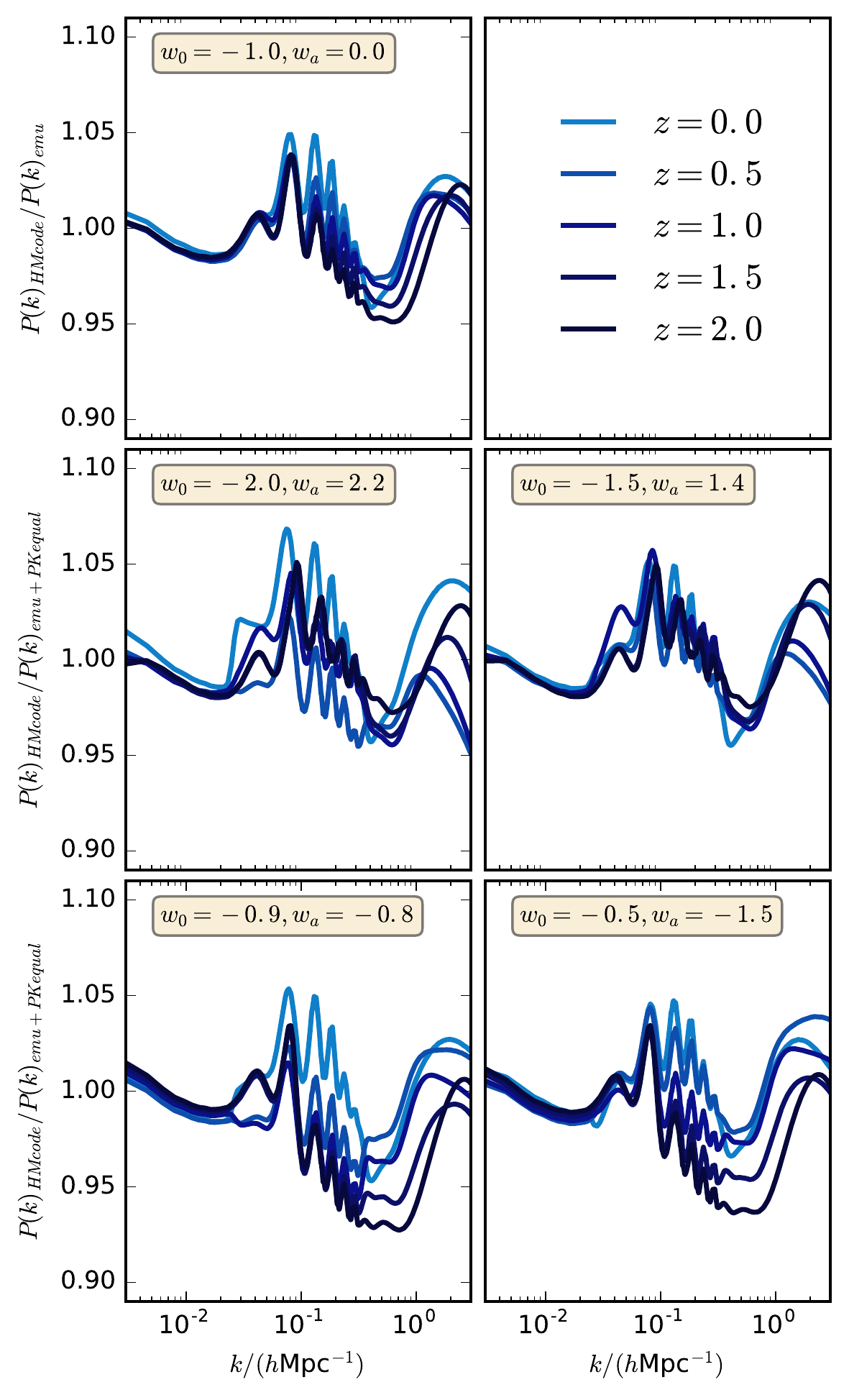}  
\caption{\label{fig:hmfig} On the top plot we show the spectral ratio
  between HMcode and COSMIC EMU, for a $\Lambda$CDM model, keeping
  mostly inside $5\, \%$ for $k<2$ and $z \leq 2$. The other plots
  refer to CPL models ($w_0$ and $w_a$ values indicated in each
  frame). In this case, COSMIC EMU is implemented by PKequal.
  Discrepancies are then somewhere wider although, apart of the case
  with $w_0 = -2.0,~w_a=2.2$, they just slightly exceed the claimed
  $5\, \%$ accuracy and only within narrow $k$ intervals.}
\end{figure} 

The improved accuracy of our algorithm can be directly appreciated if
we compare our results with the fitted Halo model provided by the
HMcode recently extended to CPL models \cite{Mead2016}
\footnote{https://github.com/alexander-mead/HMcode}.
 
In Figure \ref{fig:hmfig}, we show a comparison between the power
spectra predicted by our code PKequal applied to COSMIC EMU and the
HMcode.



As a basic example, let us consider a $\Lambda$CDM cosmology, whose {
\it equivalent $w=const$ model}, obviously, is $\Lambda$CDM itself. 
Let us then compare the results of the HMcode, for this case, 
directly with the COSMIC EMU. As expected, through the HMcode, we 
recover the results shown in Fig. 2 of \cite{Mead2015}; the 
predictions of COSMIC EMU, although slightly discrepant, are 
substantially in agreement with the HMcode at 5~\% level, as 
expected. In order to perform this comparison we considered the {\it 
node 0} set parameters indicated in Table 1 of \cite{Mead2015}: 
$\omega_b=0.0224$, $\omega_m=0.1296$, $n_s=0.97$, $h=0.72$ and 
$\sigma_8=0.8$.


Fig. \ref{fig:hmfig} let us then appreciate the performance of HMcode 
also for a number of CPL models, mostly inside the $5\%$ precision for 
$k < 2\, {\rm Mpc}^{-1}h$ and $z < 2$, apart of the case of DE with 
$w_0=-2$, $w_a=2.2\, $ at $z=0$, exhibiting a greater discrepancy in a 
short interval around $k = 0.1$~. It should be however borne in mind 
that the HMcode covers models well beyond standard DE, whose state 
equation has a CPL expression, like coupled DE models and Vainshtein 
screened modified gravity models.



\section{Conclusions} 
In this work we have first revisited the equivalence criterion 
proposed by \cite{Casarini2009}, to extend \cite{Francis2007} results 
to any redshift with no accuracy degrade. Such criterion allows us to 
obtain the spectra of models with state parameter parametrized by the 
CPL expression (target models) from suitable auxiliary models with 
constant state parameter. The auxiliary model, and hence the constant 
state parameter $w$, varies with redshift. 
 
 
This criterion, already tested in a variety of cases 
\cite{Casarini2009,Casarini2010a,Casarini2010b}, was again verified 
here for two models at the extremes of the 2--$\sigma$ likelihood 
ellipse set by the Planck experiment on the $w_0$--$w_a$ plane. 
Results were shown in detail for $w_0=-0.6$, $w_a=-1.5$, standard 
$\Omega$'s, $n_s$ and $h$ parameter values, and a set of $\sigma_8$ 
values from 0.6 to unity. 
 
Residual discrepancies were so shown to keep within $\simeq 0.5\, \%$ 
($\simeq 1\, \%$) in the range of modes $0.1<k<2$ (3) and for 
redshifts $0 \leq z \leq 3$. 
 
The remarkable precision of this equivalence points to a degeneracy of 
the imprint of DE on the non-linear matter power power spectrum. This 
is because the non-linear regime of gravitational collapse, far from 
erasing information on the past linear growth, rather carries a strong 
signature of it. Henceforth, since DE nature affects the linear growth 
of matter density perturbations in the late cosmic expansion, 
models with similar linear growth histories also show very similar 
non-linear power spectra. The conditions for the spectral equivalence 
defined by eqs.~(\ref{firstcond}) and~(\ref{secondcond}) guarantees to 
maps models with such properties. 
 
We also compared the model mapping criterion \cite{Casarini2009} with 
a few alternative ones and found that it performs best. This result 
outlines the key role of comoving distance or conformal time in 
comparing linear (and, henceforth, nonlinear) evolutions. 
 
The essential contribution of this work however concerns a set of  
new codes, based on this equivalence criterion and now made  
available. In fact, spectral equivalence allows us to provide an  
improved emulator extending the predictions of the Coyote suite to  
the non-linear matter power spectrum of DE models with CPL  
parametrization.  
 
We have also shown that the whole of the CPL parameter space 
compatible with \cite{Planck_Cosmo} cosmological constraints has 
auxiliary models in the Coyote suite, although high $\sigma_8$ values, 
marginally compatible with available cosmological limits, are just on 
the edge of the extended emulator at $z=1$, for these models where 
high $w_0$ values find their compensation in strongly negative 
$w_a$. Let us also remind that the Coyote emulator, both in the 
original and our extended versions, works up to $z=1$ at most. 
 
On the contrary, the extended Halofit expressions, also included in 
the CAMB program, formally work at any $z$ and $\sigma_8$. 
Accordingly, besides of a code that can be used individually or joined 
with Coyote Emulator, we developed a module, which can be inserted in 
the CAMB package, and allows anyone to obtain non linear power spectra 
for any $w_0$--$w_a$, in agreement with our equivalence criterion. 
This routine, e.g., can be immediately used in likelihood analyses of 
non-linear matter power spectrum measurements. 
 
\acknowledgments S.A. Bonometto acknowledges the support  
of CIFS (Inter--University Center for Space Physics), P.S.  
Corasaniti is supported by the ERC-StG ``EDECS'' no. 279954. This  
work has made use of the computing facilities of National Center for  
Supercomputing (CESUP/UFRGS), and of the Laboratory of Astroinformatics  
(IAG/USP, NAT/Unicsul), whose purchase was made possible by the  
Brazilian agency FAPESP (2009/54006-4) and the INCT-A.

\end{document}